\documentclass[onecolumn,PRL,superscriptaddress]{revtex4-2}
\usepackage[utf8]{inputenc}
\usepackage{fancyhdr}
\usepackage[T1]{fontenc}
\usepackage[margin=1in]{geometry}
\usepackage{booktabs}
\usepackage{amsmath}
\usepackage{bm}
\usepackage{bbold}
\usepackage{tabu}
\usepackage{algorithm}
\usepackage{subcaption}
\usepackage{comment}
\usepackage{algpseudocode}
\usepackage[labelfont=bf]{caption}
\usepackage{color,soul}
\usepackage{float}
\usepackage{lettrine}
\usepackage[italicdiff]{physics}
\usepackage{tikz}
\usetikzlibrary{positioning,chains,fit,shapes,calc}
\usepackage{lipsum}
\usepackage{qcircuit}
\newtheorem{prop}{Proposition}
\newtheorem{definition}{Definition}

\newtheorem{lemma}{Lemma}
\newtheorem{remark}{Remark}

\newcommand{\RS}{\textsf{RS}}
\newcommand{\ext}{\textsf{Ext}}

\newenvironment{descr*}%
  {\begin{description}%
    \setlength{\itemsep}{2pt}%
    \setlength{\parskip}{0pt}}%
  {\end{description}}

\begin{document}
\DeclareFixedFont{\elevenpt}{\encodingdefault}{\familydefault}{\seriesdefault}{\shapedefault}{11pt}
\definecolor{myblue}{RGB}{80,80,160}
\definecolor{mygreen}{RGB}{80,160,80}

\title{Flexible polar encoding for information reconciliation in QKD}

\author{Snehasis Addy}
\email{snehasis.addy@ucalgary.ca}
\affiliation{Institute for Quantum Science and Technology, University of Calgary, Canada}
\affiliation{Department of Physics and Astronomy, University of Calgary, Canada}
\author{Sabyasachi Dutta}
\affiliation{Department of Computer Science, University of Calgary, Canada}
\author{Somnath Panja}
\affiliation{Department of Computer Science, University of Calgary, Canada}
\author{Kunal Dey}
\affiliation{Department of Computer Science, University of Calgary, Canada}
\author{Reihaneh Safavi-Naini}
\affiliation{Institute for Quantum Science and Technology, University of Calgary, Canada}
\affiliation{Department of Computer Science, University of Calgary, Canada}
\author{Daniel Oblak}
\email{doblak@ucalgary.ca}
\affiliation{Institute for Quantum Science and Technology, University of Calgary, Canada}
\affiliation{Department of Physics and Astronomy, University of Calgary, Canada}

\begin{abstract}
    Quantum Key Distribution (QKD) enables two parties to establish a common secret key that is information-theoretically secure by transmitting random bits that are encoded as qubits and sent over a quantum channel, followed by classical information processing steps known as information reconciliation and key extraction. Transmission of information over a quantum channel introduces errors that are generally considered to be due to the adversary’s tampering with the quantum channel and needs to be corrected using classical communication over an (authenticated) public channel.
Commonly used error-correcting codes in the context of QKD include
cascade codes, low-density parity check (LDPC) codes, and more recently polar codes. In this work, we explore the applicability of designing a polar code encoder based on a channel reliability sequence. We show that the reliability sequence can be derived and used to design an encoder independent of the choice of decoder.  We then implement our design and evaluate its performance against previous implementations of polar code encoders for QKD as well as other typical error-correcting codes.
A key advantage of our approach is the modular design, which decouples the encoder and decoder design and allows independent optimization of each. 
Our work leads to more versatile polar code-based error reconciliation in QKD systems that would result in deployment in a  broader range of scenarios.
\end{abstract}

\maketitle

\section{Introduction}

Technologies of the second quantum revolution have the potential to overhaul and
expand the capabilities of computer and communication systems. Quantum computers can efficiently solve many computational problems that have been considered intractable for decades using classical computers, and quantum communication has enabled provably secure communication between distant parties, against an adversary with unlimited computational power. One of the first 
steps in the deployment of quantum communication networks is ensuring security for communication.

A quantum key distribution (QKD), as the name suggests, is a key distribution
protocol that aims to establish a shared key between two or more parties,  providing information-theoretic security for the communication.
A QKD protocol has two major steps - quantum state sharing and classical post-processing. In this paper, we will consider a BB84-based QKD protocol steps of which are as follows. 

In the first step, the sender (Alice) randomly encodes (classical) key bits into quantum states (quantum bits aka the qubits) prepared { in one of the two} diagonal bases (selected randomly): computational \{$\ket{0},\ket{1}$\} or Hadamard \{$\ket{+}$,$\ket{-}$\} of a two dimensional Hilbert space, where $\ket{0} = (1,0)^T$, $\ket{1} = (0,1)^T, \ket{\pm} = \frac{1}{\sqrt{2}}(\ket{0}\pm \ket{1})$. Alice sends the qubits to the receiver (Bob) who measures them in one of the two bases, again selected randomly, and stores the results.
The second step, which is the classical post-processing, is further divided into four sub-parts, viz.  {\it sifting}, {\it parameter estimation}, {\it error correction}, and {\it privacy amplification}. 

In the sifting step, Alice shares the list of all her randomly chosen preparation bases with Bob and only keep the received qubits for which Bob chose the same basis for his measurement. Since Alice and Bob each have two basis choices for preparation and measurement respectively, they will have chosen on an average 50$\%$ of the bits (see \cite{wolf2021quantum} chapter 4). The length of the resulting shared bit string is $M$.
In the parameter estimation step, the quantum bit error rate (QBER) is calculated by comparing a small subset of both Alice's and Bob's keys ($e$), which is then discarded. The remaining key $N < M$ is known as the raw key, which may be only partially correlated and partially secret. The next steps ensures that the output key is completely correlated and secret.
Error correction and privacy amplification are carried out to remove the errors introduced by the channel and reduce the entropy loss as a result of eavesdropping respectively. In QKD an eavesdropper (Eve) will always introduce errors in the shared raw-key. This is fingerprint of Eve is, indeed, the source of the provable security of QKD. In actuality, some errors will also be a result of imperfect devices such as transmitters and detectors. Errors due to either source are, however, indistinguishable i.e. the errors generated from eavesdropping and the ones generated from imperfect devices have the same effect on the key. To simplify the model and analysis while giving the adversary the maximum capability, the general assumption is, hence, that all the errors introduced are caused by Eve.

Error correction corrects errors in the raw key. The leaked information through error correction will be taken into account and removed during privacy amplification.
The first protocol for error correction in QKD (with associate information leakage)  is the Cascade protocol \cite{Cascade1994}.
This protocol was a widely used reconciliation protocol for practical implementations of QKD experiments for many years. 
{ Although simple to implement, Cascade is a highly interactive protocol and often has significant latencies. 
Further development of QKD protocols and the required hardware has significantly improved the key rate of QKD protocols \cite{li2023high,islam2017,pirandola2020} 
and, thus, latency in post-processing is a major drawback. More recent protocols use less interaction by employing error-correcting codes such as Low-Density Parity Check (LDPC) codes \cite{gallager,elkouss2009}. 
LDPC codes are known to be capacity-approaching \cite{chungLDPC}, {\em i.e.} information rate approaching Shannon limit \cite{shannon}.}
{ A new approach to error-correcting codes, introduced by Arikan \cite{arikan}, also known as Polar codes are also proven to achieve the Shannon limit. Moreover, these are known to have low-complexity encoding and decoding and, provide promising alternative for implementation.}

\vspace{0.2cm}
\noindent{\it Motivation.}
{ Polar codes are error-correcting codes based on the principle of channel polarization. Here, by channel we mean a  medium used to transmit the signal from transmitter to
receiver. The encoding process in polar codes includes channel transformation using a generator matrix to polarize the input channels into channels with high and low capacity as defined in Def. \ref{def:channel_cap}. Identification of such transformed channels makes up the code construction in polar codes. More details on this encoding procedure are discussed later in the paper.} {Polar codes are considered to be state-of-the-art error-correcting codes for certain channels e.g. control channels in 5G New Radio \cite{bioglio}.} The codes were proven to be computationally efficient and reach the Shannon limit {and therefore are considered to be a suitable replacement for Low-Density Parity-Check (LDPC) codes}. 

 In QKD, error correction has
 relied heavily on LDPC codes as in the work by D. Elkouss et al. \cite{elkouss2009} where they describe an LDPC reconciliation protocol that improves significantly on latency issues of Cascade protocol. This begs the question if a transition to Polar codes for QKD could yield similar benefits to those granted to classical communication. Multiple attempts to use polar codes for post-processing in QKD were made by a handful of research groups, giving a variety of code construction of polar codes \cite{arikan,mori2009performance,nakassis2014}. The most prominent of them is based on the work of Mori and Tanaka \cite{mori2009performance,mori2009performance1} who proposed a density evolution-based approach. 

Over the years, the approach originally proposed by Arikan based on Bhattacharyya parameters (see Def. \ref{def:Bhatta_param}) has not been widely used in practice. 
A strong reason to revisit Arikan's approach is its favorable feature of the encoder design not depending on the decoding   
procedure. 
Moreover, the calculation of Bhattacharyya parameters, albeit conceptually intricate, is computationally easy if the resultant expression for the polarized channel is known.

\vspace{0.2cm}
\noindent{\it Our Work.}
In this paper, we describe a reconciliation protocol based on polar codes. Our work mainly focuses on implementing Arikan's 
proposed approach, where he produces multiple copies of a channel and then uses the Bhattacharyya parameter \cite{arikan} to create a reliability sequence {\em i.e.} ordering of channels in terms of their channel capacity (see Def. \ref{def:channel_cap}). This allows designating good channels to carry information and bad channels as a frozen set.
We show that this construction of polar codes takes much less time compared to a construction proposed by Nakassis and Mink \cite{nakassis2014}. We implement this encoder to perform error correction for QKD. Following this we perform privacy amplification on the resultant key using an extractor. We also use our encoder for conducting { a number of} simulation experiments to measure the efficiency of polar codes, the results of this are provided in Sec. \ref{section:expt}. 
We further highlight that our encoder construction does not depend on the decoder and can be used flexibly for different block lengths. Towards this end, we first calculate the expressions of Bhattacharyya ($Z$) parameters for a binary symmetric channel (BSC) and use them in our proposed algorithm to generate a sequence that becomes useful in the encoding process. We use such an encoder and an existing decoder (viz. Successive Cancellation \cite{arikan,alamdar_SCdecoding,babar}) to carry out experiments and report the results.

\vspace{0.2cm}
\noindent{\it Paper Organization.} The rest of the paper is organized as follows. Section \ref{section:relworks} discusses previous works as well as the current status of using polar codes in QKD protocols. Section \ref{sec:prelim} (Preliminaries) includes important notations and definitions used throughout the paper.
 Section \ref{section:encoding} mainly includes the discussion on encoding in polar codes and our proposed encoding procedure along with the algorithm. The experiments completed using the proposed encoding method and their results along with a comparison with Nakassis and Mink's work \cite{nakassis2014} have been discussed in section \ref{section:expt}. In section \ref{section:fullprotocol&secanalysis} we describe the full protocol of implementing polar codes in QKD and their asymptotic as well as non-asymptotic security analysis. This is the place where we show the calculation of the secret key length. We end the paper in section \ref{section:conclusion} by summarizing all the important highlights of our paper and provide a few directions for future work.

\subsection{Related Works}
\label{section:relworks}
The first error-correcting code to be used in QKD post-processing was Cascade \cite{Cascade1994,Cascade1995} proposed in 1993. {This protocol is designed to run for a fixed number of passes. In each pass, the two parties involved namely {\it Alice} and {\it Bob} divide their respective strings into blocks of the same size. The size of the blocks is determined using the QBER calculated during the parameter estimation process and is doubled at the beginning of each pass. The main idea is to compute the parity of each block and share it with the other party. Any mismatch of parity indicates the presence of odd number of errors which can then be located using a {\em dichotomic} search -- an error found after the first pass indicates the presence of odd number of errors for previous pass. 
These errors can then be corrected by the algorithm by revisiting those passes. This makes Cascade highly interactive i.e. requires multiple back-and-forth communication rounds between the two parties \cite{elkouss2009}. 
The interactivity of Cascade introduces latency issues, thus, making it a less suitable choice for information reconciliation in high-bandwidth QKD systems. To reduce the interactivity of the Cascade protocol, the Winnow protocol, was introduced \cite{buttler2003winnow} based on Hamming codes. Although faster than Cascade, the Winnow protocol achieves lower efficiency for practical QBER values ($p<0.1$) \cite{elkouss2009}.} Subsequently, Low-Density Parity Check (LDPC) codes were introduced for QKD \cite{elkouss2009}, where Alice uses a ``parity check matrix"  to calculate a so-called syndrome of her key. This syndrome is then shared with the receiver (Bob) through a public channel and allows the decoder to reconcile the shared key of {\it Bob} with {\it Alice}. This shows the non-interactive nature of LDPC codes, which along with no latency issues makes it a popular contender for QKD. However, good LDPC codes usually have a very long code length 
\cite{sarvaghad2020new}, which causes a high computational complexity and requires large hardware memory to store the matrix at both the encoder and decoder \cite{sarvaghad2020new}. This limitation prompted continued research into more efficient error correction protocols.

Arikan \cite{arikan} introduced Polar codes as a new error-correcting code based on the principle of channel polarization. This new approach is notable for achieving the channel capacity limit and yielding a more efficient encoder and decoder, i.e., the complexity with block-length $N$ goes as ${\cal O}(N\log N)$ \cite{guruswami2020arikan}.  
These features have spurred the adoption of polar codes in the 5G New Radio control channel \cite{bioglio}. In the context of QKD, Jouguet and Kunz-Jacques \cite{jouguet} performed a comparative analysis between LDPC and polar codes and found out that polar codes feature high-speed recursive decoding and achieve CPU decoding speeds similar to LDPC GPU decoding speeds. Nakassis and Mink \cite{nakassis2014} also used polar codes for information reconciliation for a non-interactive decoder. Later, Nakassis \cite{nakassis2017} proposed an interactive polar decoder for QKD, and demonstrated it to be efficient under certain conditions, and also showed how this interactive decoder can be used to select the correct set of frozen bits. {Recently, Fang et al. \cite{nature2022} proposed a new scheme for faster post-processing using polar codes by modeling the eavesdropping activity of Eve as a wiretap channel. }{However, the details of implementation security and efficiency measurements were not provided.

 With regards to {\em privacy amplification}, {Bennett et al. \cite{bennett1988privacy} proposed an approach in the BB84 protocol \cite{BB84}, which relies on universal hash families introduced by Wegman and Carter 
\cite{carter1977universal,wegman1981new}.  Bennett, Brassard,  Crepeau, and Maurer \cite{bennett1995generalized} generalized this idea (BBCM protocol) to encompass probabilistic eavesdropping strategies. Another approach to privacy amplification, introduced a few years later \cite{maurer2000information}, employed an extractor, which is a function capable of extracting uniformly random bits from a weakly random source, along with a small number of additional random bits \cite{nisan1996randomness}.  
 Most renowned security analysis of BB84 protocol was done by Shor and Preskill \cite{Shor_2000}. 
 Security of QKD protocols along with privacy amplification was also studied in the works by  \cite{deutsch1996quantum, lo1999unconditional}. The work by Tomamichael et al. \cite{tomamichel2012tight} considers concrete security analysis under non-asymptotic cases (finite key regime).
 
\section{Preliminaries}
\label{sec:prelim}
{ 
In this section, we discuss the definitions and notation that are required for the remainder of the manuscript.

\noindent{\bf Notation.}
We will mostly follow the notation used by Arikan \cite{arikan}, but for the sake of conciseness, we describe only that used in our paper.\\
We denote the Kronecker product of two matrices (${\bf A, B}$) by ${\bf A} \otimes {\bf B}$. If ${\bf A} = {\bf B}$, we use ${\bf A}^{\otimes 2}$ as a shorthand for the Kronecker product of two identical matrices and similarly for the Kronecker product of ${\bf A}$ with itself $n$ times we use ${\bf A}^{\otimes n}$.
$X^N$ denotes a binary vector of length $N$ and $U^{i+c}=\langle u^iu^c \rangle$, where  $i$ are the indices designating information set and $c$ are the indices designating frozen set which together when concatenated make up the indices of a vector of length $N$ as denoted by $\langle u^iu^c \rangle$.\\

\begin{definition}[Binary discrete memoryless channel (B-DMC)]
    A channel $W : \mathcal{X} \rightarrow \mathcal{Y}$  with input alphabet $\mathcal{X}$, output alphabet $\mathcal{Y}$, and transition probabilities $W(y | x), x \in \mathcal{X}, y \in \mathcal{Y}$ is discrete when both the alphabets ${\cal X}$ and ${\cal Y}$ are of finite sizes and is memoryless when the current output symbol depends only on the current input symbol and not any of the previous ones. 
\end{definition}
{ A special class of B-DMC that is used in this paper is BSC (binary symmetric channel) defined below.}
%Let  $W : \mathcal{X} \rightarrow \mathcal{Y}$ denote a generic Binary discrete memoryless channel ($B-DMC$) with input alphabet $\mathcal{X}$, output alphabet $\mathcal{Y}$, and transition probabilities $W(y | x), x \in \mathcal{X}, y \in \mathcal{Y}$. The input alphabet $\mathcal{X}$ will always be $\{0,1\}$, the output alphabet and the transition probabilities may be arbitrary. We write $W^N$ to denote the channel corresponding to $N$ uses of $W$; thus, $W^N : \mathcal{X}^N \rightarrow \mathcal{Y}^N$ with $W^N(y^N_1 |x^N_1)=\prod_{i=1}^N W(y_i |x_i)$.

\begin{definition}[Binary Symmetric Channel (BSC) \cite{arikan}]\label{def:BSC}
A binary symmetric channel (BSC) is a B-DMC $W : \mathcal{X} \rightarrow \mathcal{Y}$ with $\mathcal{Y} = \{0,1\}$, $W(0|0) = W(1|1)$, and $W(1|0) = W(0|1)$. 
\end{definition}
{\noindent}{ We use BSC to model errors in the quantum channel.}

Consider a setup where Alice encodes her classical key bits into qubits and sends them to Bob through a quantum channel. Bob measures the qubits and records the measurement outcomes. The quantum bit error rate calculated by both Alice and Bob is defined as follows.
\begin{definition}[Quantum bit error rate (QBER) \cite{qber}]\label{def:QBER}
The quantum bit error rate (QBER) is defined as the ratio of the number of erroneous received bits (by Bob) to the total number of bits. The QBER can be computed using the following formula: 
$$\text{QBER} = \frac{number~of~errors}{total~number~of~bits} \times 100\%$$
\end{definition}
\noindent Note that an erroneous received bit corresponds to the receiver having detected the bit value 0(1) while the sender (Alice) had encoded the bit value 1(0).

\begin{definition}[Bhattacharyya Parameter (Z) \cite{arikan}]\label{def:Bhatta_param}
The Bhattacharyya parameter $Z$  provides an upper bound on the probability of maximum-likelihood (ML) decision error when a channel is used only once to transmit a 0 or 1. $Z$ value corresponding to a B-DMC $W$ is given by $Z(W) = \sum_{y \in \mathcal{Y}}\sqrt{W(y|0)\cdot W(y|1)}$.
\end{definition}
\noindent We use the Bhattacharyya parameter to quantify the reliability of polarized channels in terms of their capacity (see Def.~\ref{def:channel_cap}).

\begin{remark}
    For BSC with transition probability `$p$' ($i.e.$, $W(1|0) = W(0|1) = p$ and $W(0|0) = W(1|1) = 1-p$), $Z(W) = 2\sqrt{(1-p)\cdot p}$.
\end{remark}

\begin{definition}[Symmetric Capacity ($I_s(W)$) \cite{arikan}] \label{def:symm_cap}The symmetric capacity of a channel $W$ is defined as:
$$I_s(W) = \sum_{y \in \mathcal{Y}}\sum_{x \in \mathcal{X}} \frac{1}{2} W(y|x) \log \frac{W(y|x)}{\frac{1}{2} W(y|0)+ \frac{1}{2} W(y|1)}.$$
    
\end{definition}
\noindent Symmetric Capacity ($I_s(W)$) is the highest rate at which reliable communication is possible across $W$ using the inputs of $W$ with equal frequency. The symmetric capacity $I(W)$ equals the channel capacity (see Def. \ref{def:channel_cap}) when $W$ is a symmetric channel \cite{arikan}. We will use this to describe channel polarization in the next subsection.

\begin{definition}[Channel Capacity (C) \cite{cover1999elements}] \label{def:channel_cap} Suppose $X$ be the channel input symbol and $Y$ be the channel output symbol over $\mathcal{X},\mathcal{Y}$ respectively for $W : \mathcal{X} \rightarrow \mathcal{Y}$. The channel capacity is defined as the maximum of mutual information between the input and the output symbol.
$$C = \max_{p_X(x)} I(X;Y)$$
The maximum is taken over all the possible input distributions $p_X(x)$.
    
\end{definition}

\begin{definition}[($n,k$) Error Correcting Code \cite{guruswami2022}]
\label{def:ECC}An error-correcting code over an alphabet $\Sigma$ can be defined using a pair of maps (Enc, Dec), where Enc: $\Sigma^k \to \Sigma^n$ is an injective map from $k$ symbols to $n$ symbols of a coded form, and a decoding map Dec: $\Sigma^n \to \Sigma^k$ from $n$ symbols back to $k$ symbols. 
\end{definition}
\noindent Here, $n$ is the block length, and $k$ is the dimension of the code. An $(n,k)$ error correcting code is specified by a $n \times k$ generator matrix $G$ with elements $\in \Sigma^k$. For a message $x \in \Sigma^k$ the codeword $y \in \Sigma^n$ can be written as $Gx$. The dual of generator matrix $G$ is a parity check matrix $H$. For a codeword $y$ the syndrome can be written as $Hy$. Note that, although polar codes uses multiplication of the string by a generator matrix (see Sec. \ref{section:encoding}), the procedure is not the same as explained for (n,k) error correcting codes (see Def. \ref{def:ECC}). It is important to clarify that in case of Arikan's construction of polar codes, the generator matrix has dimensions $n\times n$ and the notion of syndrome, also associated with LDPC, is not relevant. This is further explained in Sec. \ref{section:encoding}.

\begin{definition}[Code Rate]
\label{def:code_rate}The code rate of an (n,k) error-correcting code is defined as 
\begin{equation}
    R = \frac{k}{n}
\end{equation}      
\end{definition}
\noindent According to Shannon's noisy channel coding theorem (see \cite{shannon}) the maximum allowed code rate for an $(n,k)$ code used to correct error over a channel is equal to its channel capacity $C$.

\begin{definition}[Frame Error Rate (FER)]
\label{def:FER}For a decoder $D$, FER is defined as the probability of failed decoding.
\end{definition}
\noindent Since the failed decoding does not lead to a reconciled key, we use this to account for the generated key per bit sent over the channel. We will use this in sec. \ref{section:expt} to calculate the average secure key established per bit.

\subsection{Channel Polarization: Definitions \& Notation}\label{subsec:2.1}
Channel polarization helps to create out of $N$ independent copies of a channel $W$ which is a B-DMC, a second set of $N$ channels ($W_N^{(i)}: 1\leq i \leq N$), showing the polarization effect such that when $N \to \infty$  the symmetric capacity $I_s(W_N^{(i)})$ (Def. \ref{def:symm_cap}) tends to either 0 or 1 for all but a vanishing fraction of indices $i$. This operation consists of two phases viz. {\em channel combining} and {\em channel splitting}.

\noindent {\bf Channel Combining:} In this phase a vector channel $W_N : \mathcal{X}^N \rightarrow \mathcal{Y}^N$, where $N$ is of the form $ 2^n$ ($n \geq 0$), is produced by combining copies of a given B-DMC, $W$, in a recursive fashion. 
The recursion begins at the 0-th level ($n = 0$) with only one copy of $W$ and we set $W_1 = W$. The general form of recursion takes two independent copies of $W_{N/2}$ and combine them to form $W_{N}$. The transition probabilities of two channels $W_N$ and $W^N$ are related by:
\begin{equation}
    W_N(y_1^N|u_1^N) = W^N(y_1^N|u_1^N G_N)
\end{equation}
for all $y_1^N \in \mathcal{Y}^N$ and $u_1^N \in \mathcal{X}^N$. Where, $G_N = B_N F^{\otimes n}$, for any $N = 2^n$, $n \geq 0$. Here $B_N$ is a suitably chosen permutation matrix also known as bit reversal and $F = $ $\begin{pmatrix}
    1 & 0 \\
    1 & 1
\end{pmatrix}$ is the kernel matrix. For more details, we refer to [\cite{arikan}, Section I-B].

\noindent {\bf Channel Splitting: } This phase splits $W_N$ back into the set of $N$-binary input coordinate channels $W_N^{(i)}: \mathcal{X} \rightarrow \mathcal{Y}^N \times \mathcal{X}^{i-1}$, $1 \leq i \leq N$, defined by the transition probabilities. 
\begin{equation}\label{eqn:chnlsplit}
    W_N^{(i)}(y_1^N,u_1^{i-1}|u_i) = \sum_{u_{i+1}^N \in \mathcal{X}^{N-i}} \frac{1}{2^{N-1}} W_N(y_1^N|u_1^N),
\end{equation}
where ($y_1^N,u_1^{i-1}$)denotes the output of $W_N^{(i)}$ and $u_i$ is its input.

\noindent {\bf Bhattacharyya parameter for polarized channels \cite{arikan}.} Using the result in Eqn. \ref{eqn:chnlsplit} and Def. \ref{def:Bhatta_param}, we can re-write the Bhattacharyya parameter $Z$ (see \cite{arikan} Eqn. (7)) as:
\begin{equation}
    Z_N^{(i)} \equiv Z(W_N^{(i)}) = \sum_{y_1^N \in \mathcal{Y}^N} \sum_{u_{1}^{i-1} \in \mathcal{X}^{i-1}} \sqrt{W_N^{(i)}(y_1^N,u_1^{i-1}|0) W_N^{(i)}(y_1^N,u_1^{i-1}|1)}.
\end{equation}
For the rest of the paper, we will use $Z_N^{(i)}$ as the shorthand notation for $Z(W_N^{(i)})$.
}

\section{Encoding algorithm of polar codes}
\label{section:encoding}
{ Arikan proposed an encoding procedure in polar code that uses a 2-bit kernel $\mathbf{F}_2 =
\begin{pmatrix}
    1 & 0 \\
    1 & 1
\end{pmatrix}$ recursively along with a suitable bit-reversal matrix $B_N$ (capturing the polarization phenomenon) to encode a $N=2^n$ bit binary 
string. The string is multiplied by the generator matrix  $\mathbf{G}_2^{\otimes n} = B_N \mathbf{F}_2^{\otimes n}$, where ${\otimes n}$ denotes the $n^{th}$ order Kronecker product of  $\mathbf{G}_2$, to output the codeword.
On a high level, the above construction of \cite{arikan} 
polarizes multiple channels to obtain a set of ``good'' channels with a near-optimal capacity of 1 and a set of ``bad'' channels that have nearly useless capacity of 0. The good channels are used to transmit the information bits and the bad ones are ``frozen'' to a known fixed value (usually 0). This phenomenon is captured by the left multiplication of $\mathbf{F}_2^{\otimes n}$ by $B_N$ which essentially rearranges the rows of $\mathbf{F}_2^{\otimes n}$.}  
{In Arikan's design of polar codes \cite{arikan}, the task of encoding includes determining an information set ${\cal A}$ of size $K \leq N$ where $Z_N^{(i)}\leq Z_N^{(j)}$ for all $i\in {\cal A}$ and $j \in {\cal A}^c$, where ${\cal A}^c$ is the complement of set ${\cal A}$. This is referred to as the {\em Reliability Sequence} (\RS). The term $Z_N^{(i)}$ denotes ``Bhattacharyya parameter'' of a channel, defined in Def. \ref{def:Bhatta_param} and further explained in Sec. \ref{subsec:2.1}.
To simplify performance analysis of polar codes (averaging over ensemble) Arikan did not specify how the set of frozen bits must be chosen and noted that ``it appears that the code performance is relatively insensitive to that choice''. In the following section we set out to explicitly compute the reliability sequence.  

\subsection{Finding the Frozen set.} 
The reliability order of the bit-channels depends on the type of channel and on the code length and poses a computational challenge when a 
range of code lengths and rates are considered (e.g. $5G$ New-Radio).
 In addition to  the Bhattacharyya parameter-based approach,
Arıkan also proposed  Monte-Carlo simulation
to estimate the reliabilities of bit-channel.
Further progress in the area has allowed the generation of a universal reliability sequence that can be used independent of the channel condition and used in 5G standardization \cite{bioglio}.

In this paper, we go back to Arikan's original way of 
 identifying good and bad channels, by calculating the reliability of channels and arranging them in the ascending/descending order of their reliability parameters, and choosing the set with the higher reliability values (smaller $Z$ values) as the information set, and determining the locations of the frozen bits and the information set.}
{Recall that Bhattacharyya parameter $Z$ (see Def. \ref{def:Bhatta_param}) corresponding to a { binary-input}  discrete memoryless channel is given by $Z(W) = \sum_{y \in \mathcal{Y}}\sqrt{W(y|0)\cdot W(y|1)}$. For a BSC with transition probability `$p$' $W(y|0)$ becomes either $1-p$ or $p$ and correspondingly $W(y|1)$ becomes either $p$ or $1-p$, thus making the Bhattacharyya parameter $Z(W) = 2\sqrt{(1-p)\cdot p}$.} We emphasize that the lower the $Z$ values, the better is the channel (in terms of capacity) \cite{arikan}.
An alternative approach to encoding polar codes was proposed by 
Mori and Tanaka \cite{mori2009performance} as the density evolution method, which was later improved by Tal and Vardy \cite{tal}, and guarantees theoretical accuracy.
%but has a 
%{ high computational cost ?? (might not be true)}.
In addition to that, Nakassis and Mink proposed a simulation-based approach, which is discussed in the next section. 
%Nakassis and Mink also proposed a simulation-based approach to converge to a set of frozen bits \cite{nakassis2014} (non-interactive). 
We observe that these implementations are decoder-dependent and their { results} might vary when implemented on different decoders.}

\subsection{Combined design of  encoder and decoder system }\label{subsec:3.2}

{ In the context of QKD, Fang et al. \cite{nature2022} proposed a reconciliation protocol using Wyner's wiretap channel model \cite{wyner}. They have used polar code construction based on Mori and Tanaka's density evolution technique \cite{mori2009performance, mori2009performance1}. In contrast to Arikan's Bhattacharyya parameter-based approach, density evolution uses decoding error probability of each channel to separate the good from the bad ones.
The set of channels for which the sum of decoding error probability is minimum is chosen to carry information bits and the rest are used for frozen bits \cite{mori2009performance,mori2009performance1}.}

On the other hand, Nakassis and Mink \cite{nakassis2014} proposed two methods to converge to the set of good and bad channels using simulations. The first method uses an initial set of Z-values and simulates the encoding and decoding algorithm and in the second approach, they provided a code construction purely based on simulations. We will briefly review both approaches in the following.\\
\textit{Z-values + Simulation approach.}
In the first approach, an initial subset of the frozen bits was chosen consisting of $J_1 = 1.2 \times N  \times h(p)$ bits with the highest Z-values (recall that the Z-value of a channel is inversely proportional to their capacity), where $N$, $p$, and $h(p)$ respectively denote the block length, crossover probability, and Shannon entropy. {(Note that the number of frozen bits required is $20\%$ over the Shannon limit to take into account the imperfections of the error-correcting code)}. For simulation, the decoder failure rate {\em i.e.} the ratio of the number of times the decoder fails to recover the codeword to the total number of decoding attempts, was chosen to be at most 10\%  and 500 simulations (encoding-decoding) were run with QBER (here we denote it as `$p$'). A record of all the errors occurring in the decoder was kept in a histogram.
If after decoding, the failure rate $\leq 10\%$ then the set of frozen bits is not modified. Else, the set of frozen bits is expanded by selecting up to $J_2 = 0.1 \times  N \times h(p)$ ({ another $10\%$ added above Shannon limit}) bits where errors occur and restart the experiment. If the number of positions in error is greater than $J_2$, then the $J_2$ positions with the highest histogram counts are chosen.\\
\textit{Simulation-only approach.}
In this approach, instead of using the Z-values for finding the initial subset of frozen bits, the authors use 1000 preliminary data sets { (chosen randomly)} to acquire the initial subset (for more information see \cite{nakassis2014}). The rest of the steps are the same as before. 
The authors reported that the second (purely) simulation-based approach yielded better results in terms of code efficiency. We note that in both cases, \cite{nakassis2014} used a successive cancellation (SC) decoder to implement their polar code decoder. 

{ From the above description of the method, we see that Nakassis and Mink's code construction take multiple rounds of simulation to produce a set of information and frozen bits. In addition to that, each round of simulation also requires the involvement of a decoder, making the encoding process a resource-intensive job compared to Arikan's method based on $Z$ parameter. Using this encoding method the authors of \cite{nakassis2014} also proposed a key-establishment protocol described in \ref{appendix:B}.}

\subsection{Our encoder}
    It is known that error in the quantum channel can be modeled by a BSC. Here we aim to design a flexible encoder that can be used in the protocol (described in \ref{appendix:B}) of \cite{nakassis2014}, in the sense that, $(i)$ it is designed independently of the decoder, such that potential improved decoders can be employed without impact on the encoding, and $(ii)$ it works with different block lengths as required by the channel error and QKD design. We compare our encoder with that of \cite{nakassis2014} using the same decoder, i.e., the SC decoder. For more information on the SC decoder, we refer to \cite{yuan-parhi}. The polar encoder in Nakassis and Mink \cite{nakassis2014} is { designed} for specific lengths and decoder, and is computationally expensive { since it relies on multiple rounds of simulation}.
 { We follow Arikan's proposal to polarize $N$ independent and identical BSC channels to output $N$ polarized channels. {The effect of polarization here is captured by the computation of the Bhattacharyya parameter $(Z)$. The channels are transformed in the following manner.}

The transformation starts with $N$ copies of $W_1^{(1)}$. After the first step, they are transformed into $N/2$ copies of ($W_2^{(1)},W_2^{(2)}$) which following the next iteration admits $N/4$ copies of ($W_4^{(1)},W_4^{(2)},W_4^{(3)},W_4^{(4)}$). It continues to form a tree pattern, each representing a transformation of the form ($W_{2^i}^{(j)},W_{2^i}^{(j)}$) $\to$ ($W_{2^{i+1}}^{(2j-1)},W_{2^{i+1}}^{(2j)}$).  This at local level becomes single-step channel transformation shown in Fig \ref{fig:tree}. At the leftmost node of the tree in Fig. \ref{fig:tree} (parent node) there are $N$ independent copies of $W_1^{(1)}$, in the next level (children) there are $N/2$ independent copies of $W_2^{(1)}$ and $W_2^{(2)}$; the pattern goes on for all the other levels. Therefore, with each step moving from left to right, the number of channel types gets doubled but the number of independent copies gets halved. Our proposed channel transformation is captured by the following proposition.
  
\begin{prop}
For any single-bit BSC channel ($W$), $N = 2^n$, $n \geq 0$, $1 \leq i \leq N$, for the transformation ($W_N^{(i)},W_N^{(i)}$) $\to$ ($W_{2N}^{(2i-1)},W_{2N}^{(2i)}$), 
the expressions for $Z(W_{2N}^{(2i-1)})$ and $Z(W_{2N}^{(2i)})$ can be calculated as follows: 
\begin{align}
\label{eqn:BP for BSC}
    &Z(W_{2N}^{(2i-1)}) = Z(W_{N}^{(i)})\sqrt{2-{Z(W_{N}^{(i)})}^2} \\
   &Z(W_{2N}^{(2i})) = Z(W_{N}^{(i)})^2.
\end{align}
where $Z(W_1^{(1)}) = 2\sqrt{(1-p)\cdot p}$.
\label{prop1}
\end{prop}
The proof of Proposition \ref{prop1} is provided in \ref{appendix:A}. Using expressions recursively ($n$ times) we find the Bhattacharyya parameter for length $N =2^n$ codes, which can then be arranged in ascending/ descending order to output the \RS~(channels with $Z(W_{N}^{(i)})$ near $0$ are used to send information bits) for a BSC based encoder. 
 This allows the polar encoder to be designed independent of the choice of a decoder, and be computationally efficient.}  An \RS~for BECs is available only for block length $N = 2^{10}$, published by ETSI \cite{specification2018ts} and to the best of our knowledge, this is the only known list of an \RS~available publicly; no such \RS~for BSC is known. In this paper, we provide an algorithm (\ref{alg:rs}) to find the \RS~for BSC using the $Z-$ parameter expressions given in Proposition \ref{prop1}. Our algorithm to find \RS~can also be adopted for other binary input-output DMC by suitably modifying expressions in Proposition \ref{prop1} and using it in our algorithm.  When tested we found out that the overlap of the \RS~generated by our algorithm with the one provided by ETSI (true \RS) \cite{specification2018ts} is high. This comparison was done for a \RS~of a BEC for block length 1024, shown in Fig. \ref{fig:RS}.
 Prior to our work, an algorithm for \RS~was given by Ghasemi and Uchôa-Filho \cite{ghasemi}, designed for BEC but is not suitable for BSC.

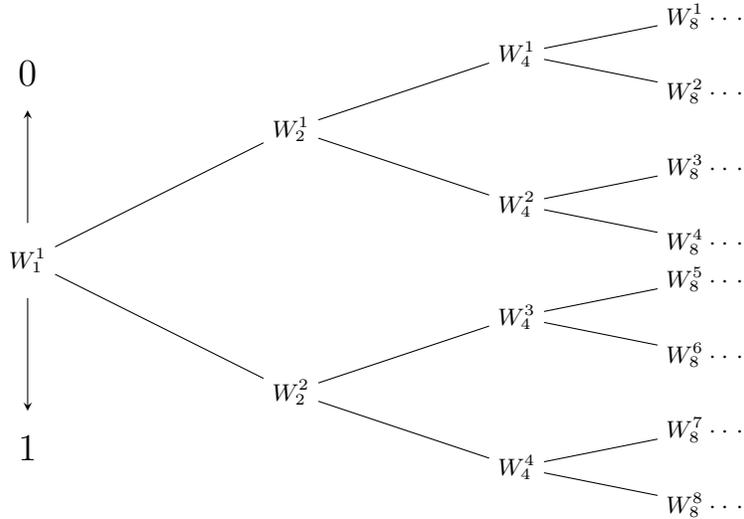
\begin{figure}[t]
\centering
\begin{tikzpicture}
	
\tikzstyle{level 1}=[level distance=3.5cm, sibling distance=3.5cm]
\tikzstyle{level 2}=[level distance=3cm, sibling distance=2cm]
\tikzstyle{level 3}=[level distance=2.5cm, sibling distance=1cm]

\node {$W_1^1$} [grow =right,sloped]
	child {node {$W_2^2$} 
	child {node {$W_4^4$}
	child{node{$W_8^8 \cdot \cdot ~\cdot$}}
	child {node {$W_8^7 \cdot \cdot ~\cdot$}}}
	child {node {$W_4^3$}
	child {node {$W_8^6 \cdot \cdot ~\cdot$}}
	child {node {$W_8^5 \cdot \cdot ~\cdot$}}}
	edge from parent [above]}
	child {node {$W_2^1$} 
	child {node {$W_4^2$}
	child {node {$W_8^4 \cdot \cdot ~\cdot$}}
	child {node{$W_8^3 \cdot \cdot ~\cdot$}}}
	child {node {$W_4^1$}
	child {node{$W_8^2 \cdot \cdot ~\cdot$}}
	child {node{$W_8^1 \cdot \cdot ~\cdot$}}}
	edge from parent [below]};

 \draw[fill= black, draw = black, {-stealth[scale=2.5]}] (0,0.5) -> (0,2);
\draw[fill= black, draw = black, {-stealth[scale=2.5]}] (0,-0.5) -> (0,-2);
\node[] at (0,2.5) {\bf {{\Large{$0$}}}};
\node[] at (0,-2.5) {\bf {{\Large{$1$}}}};
\end{tikzpicture}	
\caption{Tree representation for recursive channel transformation}
\label{fig:tree}
\end{figure}

\begin{figure}[H]
     \centering
     \includegraphics[width = 0.65\textwidth]{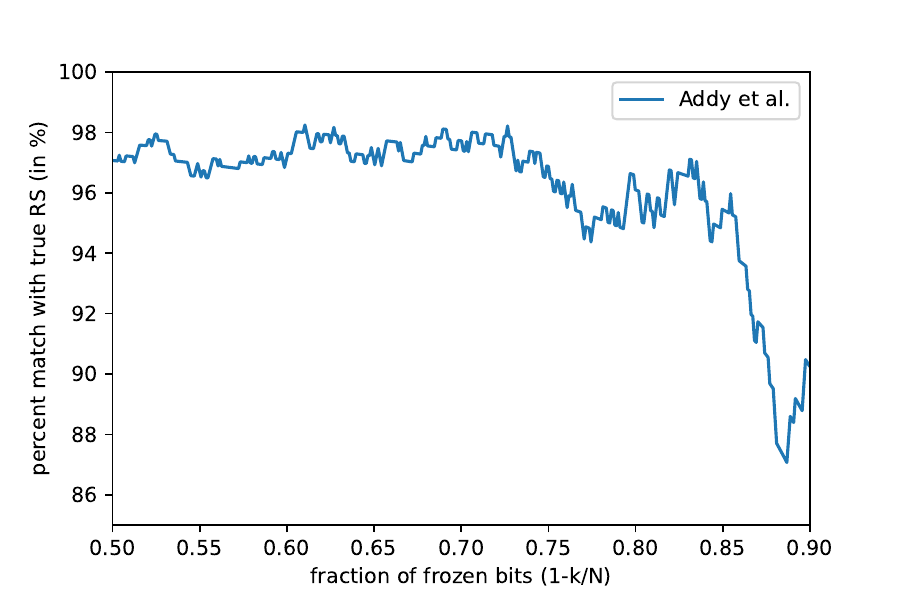}
     \caption{Overlap of \RS~generated using our algorithm (see algo. \ref{alg:rs}) with the true \RS~\cite{specification2018ts} for different fractions of frozen bits. This comparison is done for a BEC with block length of 1024.}
     \label{fig:RS}
 \end{figure}

 \subsection{Algorithm}
 In proposition \ref{prop1}, we have calculated the Bhattacharyya parameter expressions of polarized BSC channels (see Eqn. \ref{eqn:BP for BSC}). Using that we have provided an algorithm to calculate the \RS~.
 Our algorithm \ref{alg:rs} for \RS~takes values $N$ and $Z(W_2^{(1)}), Z(W_2^{(2)})$ as inputs and generates the reliability sequence as a list. The algorithm manages a list named $A$ and a list of lists named $B$ side-by-side. The first step is done manually where input values $Z(W_2^{(1)})$ and $ Z(W_2^{(2)})$ are stored in list $A$ as entry $0$ and $1$ respectively. Corresponding to this in $B$, `$1$' is stored as list 0 and `$0$' as list 1. For the rest of the cases, a loop (completing $2^{\log_2(N)}-2$ iterations) is used to fill up the remaining entries. Inside the loop, in each iteration, two entries of $A$ and $B$ are filled. We start by using a variable ($k$) with a value $k= 2$ (next empty slot in list $A$). For the first entry take the list $k-2$ of $B$, append 1 to the list, and put it as list $k$ in $B$. Similarly, calculate the value $A[k-2]\sqrt{2-{{A[k-2]}^2}}$ and append it to list $A$. For the second entry, it takes the list $k-2$ of $B$, append 0, and put it as list $k+1$ in $B$. Then, calculates the value ${A[k-2]}^2$ and append it to list $A$. At the end of the iteration, it increases the value of $k$ by $1$. Since the reliability sequence only includes the final transformed channels (not the intermediate ones), only those lists in $B$ are selected whose $size = \log_2(N)$ and a copy of those entries is made in $D$. Corresponding entries of $A$ are taken and written in $C$. Next, the sorting of list $C$ and list of lists $D$ is done using the merge sort algorithm. Here sorting is done based on values stored in $C$ and the same steps are copied for $D$. In the last step, reliability numbers are assigned as per the data stored in $D$. For that, the data stored in the list of lists $D$ from 
the beginning is considered to be in binary form and a decimal equivalent is calculated by running two loops shown in the algorithm \ref{alg:rs}. The calculated decimal equivalent is then appended in the list called $reliability$, which after completion of the loops will become the `reliability sequence'. This algorithm calculates the reliability sequence with the computational complexity of $\mathcal{O}(N \log N)$.

\begin{algorithm}[H]
\caption{Reliability Sequence}
\label{alg:rs}
\begin{algorithmic}
\Require $N$, $Z(W_2^{(1)})$ , $Z(W_2^{(2)})$
\Ensure reliability sequence
\State $deg \gets log_2(N)$, { $total \gets 2^{deg+1}-4$}, and $k \gets 2$
\State Initialize $A, D$, reliability = list [~] \Comment{Initialise as list}
\State Initialize $B, C$ = list [list[1],...,list[i],...] \Comment{Initialise as list of lists}
\State $A.append(Z(W_2^{(1)})) $, $A.append(Z(W_2^{(2)})$
\State $B[0] \gets 1$, and $B[1] \gets 0$
\For{i= 0: $\frac{total}{2}$-1}
\State Initialize $v$ =list[~] 
\State $v\gets B[k-2]$, $v.append (1)$, and $B.append (v)$
\State $A.append (A[k-2]\sqrt{2-{{A[k-2]}^2}})$
\State Initialize $w$ =list[~] 
\State $w$ $\gets B[k-2]$, $w.append (0)$, and $B.append (w)$
\State $A.append ({A[k-2]}^2)$

\State k++
\EndFor
\If {B[i].size == deg}
\State $D \gets B[i]$
\State $C \gets A[i]$
\EndIf
\State \textbf{mergesort} ($C,D$) \Comment{Sort list C increasing order and D same as C}
\For{i in D.size}
\State Initialize s= 0
\For {j = $deg-1$: 0}
\State s $\gets$ s$ +D[i,j]\times 2^{deg-j-1}$
\EndFor
\State $reliability.append (s)$ \Comment{`$reliability$' stores the reliability sequence.}
\EndFor
\label{algo}
\end{algorithmic}
\end{algorithm}

\section{Experiments}
\label{section:expt}
For the experiments, we use the \RS~obtained above to identify frozen bits and use $\mathbf{G}_2^{\otimes n}$ for encoding such that, codeword becomes $({\tt msg} \cdot \mathbf{G}_2^{\otimes n})$, where ${\tt msg}$ denotes the $N$-bit message string which includes information bits as well as frozen bits. In the experiments, we do a simulation of encoding and decoding using polar codes. For encoding we use the \RS~from our algorithm and use Arikan's encoding criteria \cite{arikan}, which gives us the codeword. To emulate the error in the quantum channel we do a probabilistic bit-flip on each bit with a probability $p$. Once we introduce errors into the codeword we then use the SC decoding technique \cite{arikan} to complete the decoding task. Throughout the experiment, we vary parameters like code rate, block length, and error rates $p$ to analyze the overall performance of polar codes. More specifically we perform the following experiments.
 \subsection{Experiment 1: Performance of discussed encoder used with SC decoding}
 In this experiment, we find the qubit error rate (QBER) that can be corrected using the proposed encoding technique given a certain code rate $K/N$, where $K$ is the number of information bits, i.e., the number of good channels from which the final key material can be extracted. Here we have kept the code rate as the independent variable. 
 { We calculate the corresponding QBER plotted on the y-axis of Fig. \ref{fig:sec_code_rate} as the maximum value of $p$ that can be reconciled by the decoder such that the FER (see Def. \ref{def:FER}) is just below or equal to the maximum allowed limit.} This process of finding the maximum QBER is shown in Fig. \ref{fig:QBER_calc}. 
 Thus, the total yield after error correction is given by $(1-FER)\times K/N$. The experiment assumes a maximum allowed FER of $5 \%$.

 \begin{figure}
     \centering
     \includegraphics[width=0.65\textwidth]{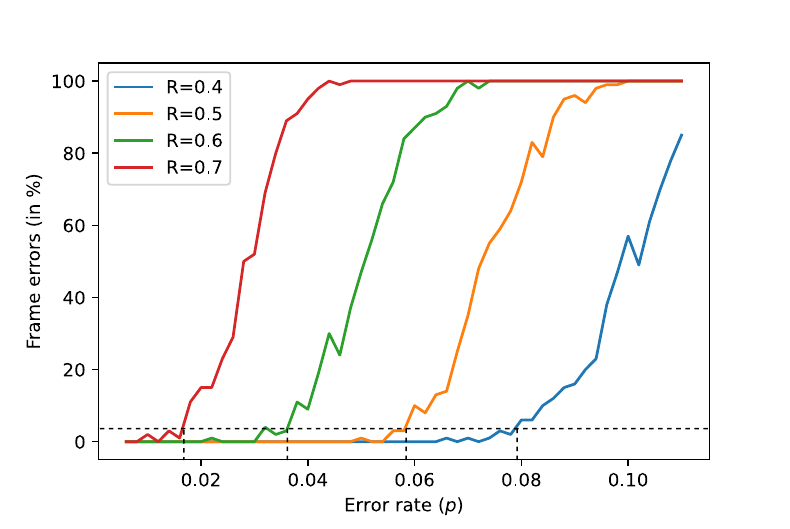}
     \caption{The graph shows the variation frame errors after error correction for different code rates ($R$) with the error rate in the channel ($p$). The dotted black line denotes the maximum allowed frame error and its intersection with the different graphs gives the maximum value of error that can be corrected corresponding to a particular code rate.}
     \label{fig:QBER_calc}
 \end{figure}
 
{\em (Observation)} Fig.~\ref{fig:sec_code_rate}$a$ shows the plot of the number of errors that can be corrected for various code rates. We can see that as the code rate decreases more errors can be corrected as, as expected, the code rate gets closer to the Shannon bound with increasing block-length $N$. 
{ For an isolated case of $N= 2^{18}$, we see that below a particular code rate and for higher QBER values the performance of the encoder becomes worse than $2^{12}\leq N \leq 2^{16}$ values (see the black curve in Fig \ref{fig:sec_code_rate}$a$). The overall shape of the plot is scattered due to the limited number of runs which is $\approx 50$ per data point and random variations.}

We use our experimental data to calculate the secrecy content per bit processed ($\gamma$) as defined in \cite{nakassis2017} given by the following expression.
\begin{equation}
    \gamma = (1-\text{FER})(\frac{K}{N} -h_2(p))
\end{equation}where $h_2(p)$ is the Shannon entropy of a binary random variable with bias $p$. This quantity corresponds to the per-bit yield of the error correction minus $h_2(p)$, which accounts for the information leaked from the quantum channel by assuming that all errors are introduced by the adversary's eavesdropping, resulting in information leakage over the quantum channel as $Nh_2(p)$ \cite{lo2005k,renner2005information}. Fig. \ref{fig:sec_code_rate}$b$ shows the QBER  that can be corrected as a function of the secrecy content per bit processed. This must be subtracted from the number of information bits sent over the classical channel $K$, giving us the final number of transmitted secret bits as: \begin{equation}
    {\gamma}{N} = (1-\text{FER})(K -Nh_2(p))
\end{equation}}
{\em (Observation) }{In Fig. \ref{fig:sec_code_rate}$b$  the variation of the entropy of the established raw key of Alice and Bob for different values of QBER and $N$ is captured.}
\begin{figure}[t]
    \centering
    \begin{subfigure}{0.45\textwidth}
        
        \includegraphics[width=1.1\linewidth]{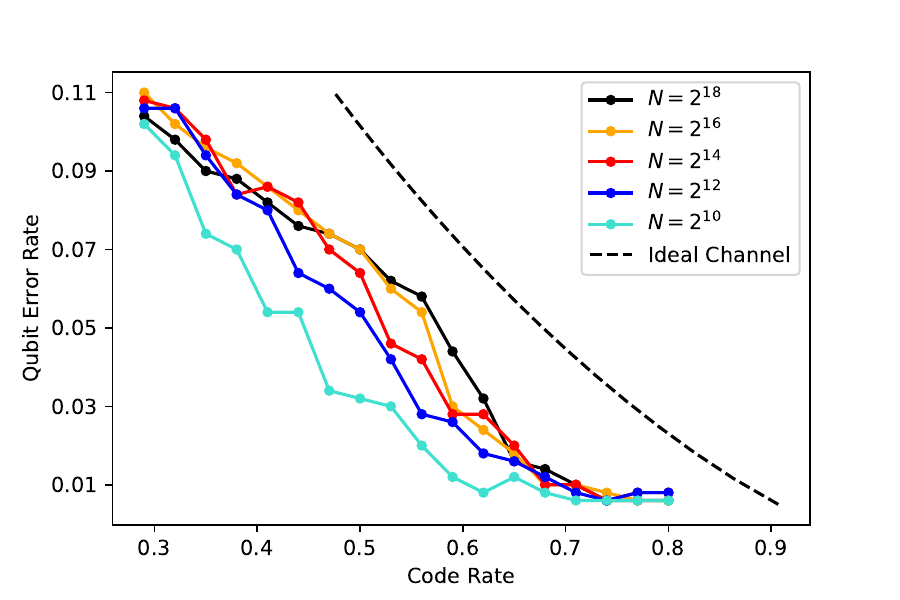} 
        \caption{}
    \end{subfigure}
    \begin{subfigure}{0.45\textwidth}
        
        \includegraphics[width=1.1\linewidth]{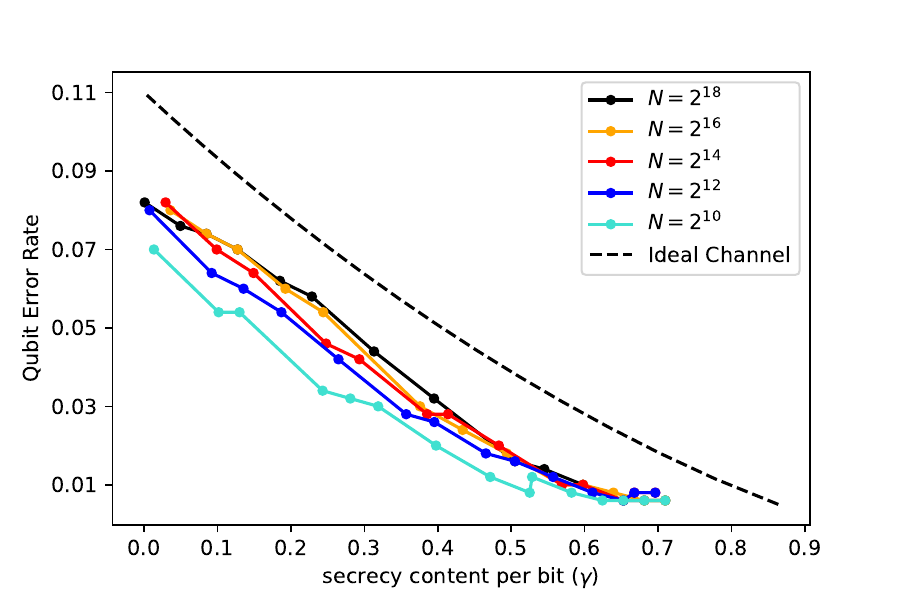} 
        \caption{}
    \end{subfigure}
    
    \caption{a) QBER vs Code rate for our implementation. Experimental Code rate for every value of QBER; 50 simulations were performed for each point. b) QBER vs Secrecy content per bit processed ($\gamma$) for our implementation, 
    where $\gamma = (1-\text{FER})(\frac{K}{N} -h_2(p))$. The FER for both plots was kept at 0.05.}
    \label{fig:sec_code_rate}
\end{figure}

\begin{figure}[t]
    \centering
    \begin{subfigure}{0.45\textwidth}
        \includegraphics[width=1.1\columnwidth]{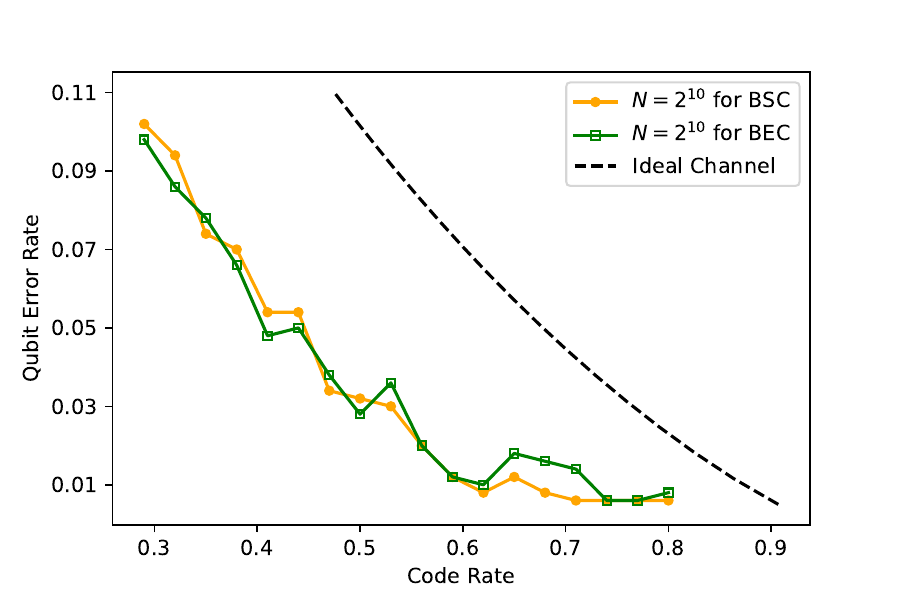} 
        \caption{}
    \end{subfigure}
    \begin{subfigure}{0.45\textwidth}
        \includegraphics[width=1.1\columnwidth]{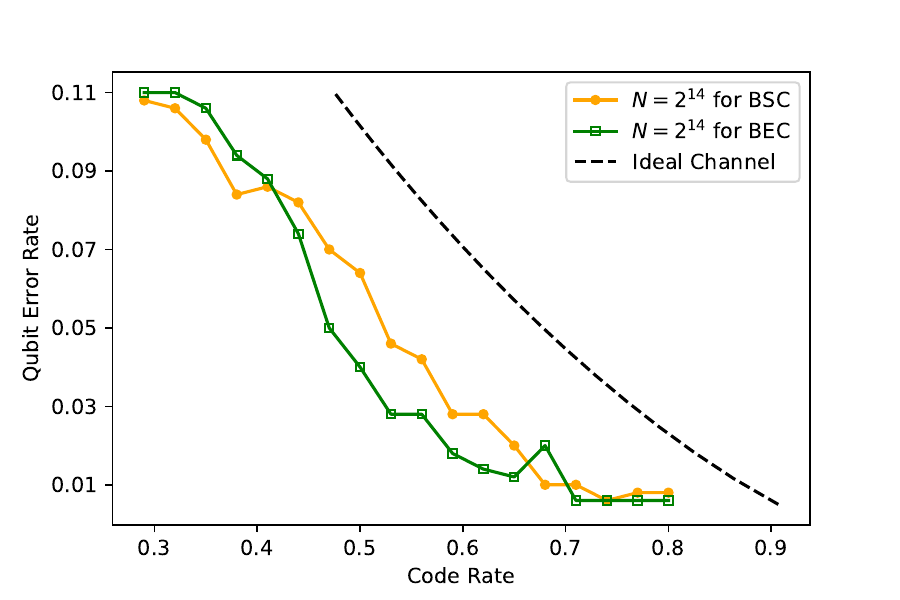} 
        \caption{}
    \end{subfigure}
    \begin{subfigure}{0.45\textwidth}
        \includegraphics[width=1.1\columnwidth]{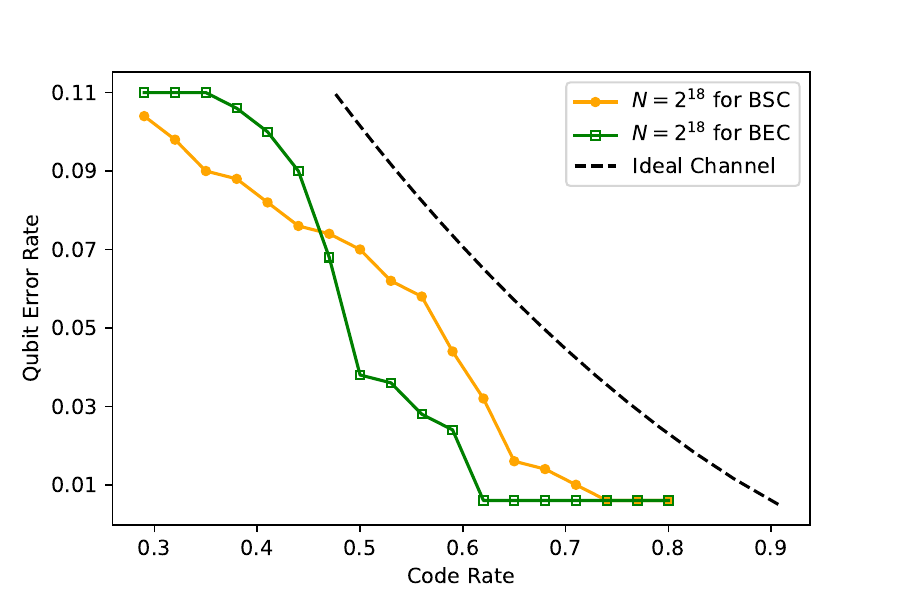}
        \caption{}
    \end{subfigure}
    
    \caption{Variation of QBER and code rate for encoders designed with \RS~of BSC and BEC for a) $N =2^{10}$, b)$N= 2^{14}$, and c) $N= 2^{18}$ respectively. The FER for all three panels was set to 0.05.}
    \label{fig:BEC-BSC comparison}
\end{figure}
\subsection{Experiment 2: Performance of encoder with \RS s for different channels types}
\label{sec:expt2}
In this experiment, we compared the performance of encoders when \RS~is derived using $Z-$ parameters, (1) assuming the channel is a BSC, (2) or a BEC with output set $\{ 0, 1, \bot\}$. The $ Z-$ parameters for the BEC are:
\begin{align}
    &Z(W_{2N}^{(2i-1)}) = 2Z(W_{N}^{(i)}) - Z(W_{N}^{(i)})^2 \\
   &Z(W_{2N}^{(2i})) = Z(W_{N}^{(i)})^2.
\end{align} for more information please see \cite{arikan}. The encoding is done as described in the paper before but with different $Z-$ parameters and decoding is done using an SC decoder. For values of $N = 2^{10}$, $2^{14}$, and $2^{18}$, code rate vs QBER is plotted in fig \ref{fig:BEC-BSC comparison}. 

{\em (Observation)} Comparing the plots in Fig. \ref{fig:BEC-BSC comparison}$a$, shows that for $N =2^{10}$, the performance of the encoder using the two \RS's (for BEC or BSC) is  similar, but if block lengths are taken larger than $N = 2^{10}$, there is a noticeable difference between the two. This difference increases when the value of $N$ is increased. This pattern can be observed by looking at fig \ref{fig:BEC-BSC comparison}$a,b,c$ simultaneously.

Another notable observation here is, for practical QBER values $(0.01-0.05)$, the performance of the BSC encoder is significantly better, $i.e.$ closer to the Shannon limit (dashed black line). However, for higher QBER values ($0.07$ or higher) BEC encoder has a better performance. It can also be observed that the two curves intersect when QBER values lie between $0.07-0.09$. 
{ We cannot explain this behavior.}

Based on our experiments we observe that the \RS~which is obtained by using $Z-$ values of BSC performs better compared to the \RS~obtained using $Z-$ values of BEC when the underlying channel is BSC and practical QBER values {\em i.e.} an error rate of $0.01-0.05$ are considered. So while computing the \RS~for polar encoding one must also take into account the nature of the underlying channel.
 
\subsection{ Comparison with Nakassis $\&$ Mink \cite{nakassis2014}.}
We summarize our results in the LHS of Table~\ref{tab:Our_result} and report Nakassis and Mink's simulation-only results in the RHS of the same Table.
\begin{table}[t] 
\begin{tabular}{ |p{1.7cm}|p{1.6cm}|p{1.5cm}| p{1.7cm}|  }
\hline
\multicolumn{4}{|c|}{Our results, $N=2^{16}$} \\
\hline
QBER ($p$)& FER& $\beta$ & Avg yield \\
\hline
0.018& 0.05& 74.71$\%$& 61.75$\%$ \\
\hline
0.024& 0.05& 74.1$\%$& 58.9$\%$ \\
\hline
0.03& 0.05&  73.2$\%$& 56.05$\%$ \\
\hline
0.054& 0.05&  80.3$\%$& 53.2$\%$ \\
\hline
0.06& 0.05&  78.8$\%$& 50.35$\%$ \\
\hline
\end{tabular}
\hfill
\begin{tabular}{ |p{1.7cm}|p{1.6cm}|p{1.6cm}|p{1.5cm}| p{1.7cm}|  }
\hline
\multicolumn{5}{|c|}{Nakassis and Mink's results $N=2^{16}$} \\
\hline
QBER ($p$)& FER lim& FER& $\beta$&  Avg yield \\
\hline
0.02& 0.1& 0.073& 93$\%$&  74$\%$ \\
\hline
0.02& 0.02& 0.013& 91.9$\%$& 77.8$\%$ \\
\hline
0.04& 0.1& 0.071& 90.1$\%$&  63.4$\%$ \\
\hline
0.04& 0.02& 0.015&  88.7$\%$& 66.2$\%$ \\
\hline
0.06& 0.1& 0.068& 87.8$\%$ & 55$\%$ \\
\hline
0.06& 0.02& 0.013& 86.2$\%$&  57.2$\%$ \\
\hline
\end{tabular}
\caption{Comparison of our results with Nakassis and Mink's simulation only method \cite{nakassis2014} for $N=2^{16}$, with $\beta =\frac{K}{N(1-h(p))}$, and Avg yield= $(1-FER)\times K/N$}
    \label{tab:Our_result}
\end{table}

We note that \cite{nakassis2014} achieve higher $\beta$-values and average yield compared to our results.
However, the computation cost of encoder design is significant. For $N = 2^{16}$, \cite{nakassis2014} takes  
$\sim468 s$ to converge to the set of frozen bits whereas
our encoder design, for the same value of $N$, we computed \RS~in $\sim10 s$ on a personal computer. The computer specifications on which the calculation of \RS~ was done are as follows: {\bf CPU:} Apple M1, {\em cores/threads:} 8/8, {\em clock speed:} 2064 - 3220 MHz, {\em RAM:} 8GB;  {\bf GPU:} Apple M1 GPU, {\em cores:} 7, {\em core speed:} 1278 MHz.  Fig. \ref{fig:sec_code_rate} shows our results, and is closely aligned with the theoretical results on polar codes. We believe our proposed approach will be very attractive in settings where multiple keys must be established e.g. by a server with multiple clients. 

\section{Full Protocol and its finite key analysis}
\label{section:fullprotocol&secanalysis}
In this section, we consider the QKD protocol and the associated finite length key rate analysis in \cite{tomamichel2012tight}, when a  polar code is used for error correction, similar to the work in \cite{nakassis2014}.

\subsection{Description of key-exchange protocol.}
\begin{figure}[t]

    \includegraphics[width =0.75\textwidth]{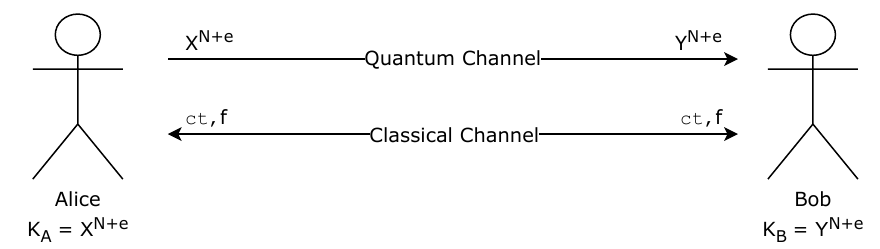}
    \hspace*{2mm}
    \framebox[\textwidth][c]{
			\begin{minipage}{0.8\textwidth}
   
{\bf Alice's Lab} (Input: Key $X^{N+e}$)
\begin{enumerate}
   \item {\em Quantum Communication}: Alice sends through quantum channel $X^{N+e}$ and Bob receives $Y^{N+e}$ \footnotemark. 
    Here $e$ number of bits are used to do parameter estimation (calculate QBER).
    \item Generate a hash function $f: \{0,1\}^N \to \{0,1\}^{\ell}$ from ${\cal F}$ based on the distribution $p_{\cal F}$ (see Def. \ref{def:two-universal-hash}). This will be used as a $(\delta,\epsilon)$-strong quantum proof randomness extractor.
    \item Generate $k$-bit random string. Place it corresponding to information (unfrozen) channels ($u_i$). The rest of the $N-k$ places are filled with frozen bits ($u_c=0$). Call $u^N= \langle u^i u^c \rangle$.
    \item Run (Polar) encoding to obtain $W$ from $u^N$.
    \item Compute: ${\tt ct} = W \oplus X^N$.
    
    \item Compute $(K_{Alice})$ = $f(X^N)$. 
    \item Send $({\tt ct},f)$ to Bob through Classical Authenticated Channel.
\end{enumerate}

{\bf Bob's Lab} (Inputs: received key $Y^{N+e}$, and $({\tt ct},f)$)
\begin{enumerate}
    \item Compute: $Z = {\tt ct} \oplus Y^N$
   
    \item Run (polar) decoder on $Z$ to get $\tilde{u}^N =  \langle \tilde{u}_i u_c \rangle$.
    \item Use encoding process to obtain $\widetilde{W}$ from $\tilde{u}^N$.
    \item Compute $\widetilde{X}^N=\widetilde{W} \oplus {\tt ct}$.
    \item Output the extracted key $(K_{Bob})$ = $f(\widetilde{X}^N)$.
\end{enumerate}
   
   \end{minipage}}
    \caption{Full key-exchange protocol}
    \label{fig:full_protocol}
\end{figure}

Fig. \ref{fig:full_protocol} shows the key exchange protocol using polar codes as an error correction code for QKD.
\footnotetext{In a practical setting Bob receives only a small subset of Alice's originally sent key. This happens due to either detector inefficiencies or loss of photons in the channel. In the step of parameter estimation, Alice and Bob together choose only to consider the qubits that are detected by Bob while the rest are ignored.}

\subsection{Universal Security Definitions.}

\noindent{\em Correctness of the protocol.}
We say that a QKD protocol is $\epsilon_{cor}$ correct if at the end of the protocol $Pr[K_{Alice} \neq K_{Bob}] \leq \epsilon_{cor}$.
Since the probability that our polar decoder can correctly decode the data is at least $1-FER$, we see that $K_{Alice} = K_{Bob}$ is true with probability at least $1-FER$. Hence in our case $\epsilon_{cor} = FER$.

\noindent{\em Secrecy of the protocol.} Following the work of \cite{tomamichel2012tight} a key $K_A$ is said to be $\epsilon_{sec}$- secret if for any classical-quantum state $\rho_{K_A,E}$ which is the composite system of Alice and Eve after a successful QKD protocol, satisfies
\begin{equation}
\label{eqn:secofQKD}
     \frac{1}{2} \norm{\rho_{K_{AE}}-\rho_U \otimes \rho_E} \leq \epsilon_{sec}.
\end{equation}
where $\rho_U = \sum_{u \in \mathcal{S}} \frac{1}{|\mathcal{S}|} \ket{u}\bra{u}$, and $\mathcal{S}$ is the key-space.

The complete security of the protocol comes from combining the correctness and the secrecy conditions. If a protocol is $\epsilon_{cor}$-correct and $\epsilon_{sec}$- secret then it is $\epsilon \geq \epsilon_{cor}+\epsilon_{sec}$ secure \cite{tomamichel2012tight}. 

\subsection{Secret key length for a finite key regime}

{ Tomamichael et al. in their work \cite{tomamichel2012tight} showed that the secret key length of BB84 protocol under the finite length analysis for can be written as:
\begin{align}
    \label{eqn:finite_key_len}
    \ell &\leq  H^{\epsilon'}_{min}(K_A|E)- leak_{EC} - \log{\frac{2}{\epsilon_{sec}^2 \epsilon_{cor}}} \nonumber \\
    &\leq N(q-h_2(Q_{max} +\mu))- leak_{EC} - \log{\frac{2}{\epsilon_{sec}^2 \epsilon_{cor}}}
\end{align}
Here, $q= \log \frac{1}{c}$ is the preparation quality of the source and is $1$ when the basis states are prepared in diagonal bases for example a theoretical BB84 model. The quantity $\mu$ can be written as $\mu = \sqrt{\frac{(e+1)(N+e)}{e^2 N} \log\frac{1}{\epsilon'}}$, where $\epsilon'$ is a parameter used to calculate the smooth conditional min-entropy (see Def. \ref{def:smooth-min}).
In Eqn. \ref{eqn:finite_key_len}, the term $N(q-h_2(Q_{max} +\mu))$ is the lower bound on $H^{\epsilon'}_{min}(K_A|E)$ which characterizes the amount of information that Eve can extract from Alice's classical bit string $K_A$. Note that Eve holds a  quantum system correlated with $K_A$. The term $leak$$_{EC}$ indicates the leakage due to error correction. To ensure the correctness of the error correction step Alice and Bob agree on a two-universal hash function (see Def. \ref{def:two-universal-hash}) through which the leakage is $\log \frac{2}{\epsilon_{cor}}$ (see \ref{appendix:Finite_key_analysis}). Following error correction Alice and Bob performs privacy amplification to determine their final key which is $\epsilon_{sec}$-secure.  
The secret key length ($\ell$) calculated is from the quantum leftover hash lemma (see lemma \ref{lemma:QLHL}). 
More details are discussed in \ref{appendix:Finite_key_analysis} and full description can be found in \cite{tomamichel2012tight}.}

For error correction, we use an $(N,k)$ polar code so can write leak$_{EC} = N-k$. We will take $\epsilon_{cor}= FER = 0.05$, $\epsilon_{sec}=0.5\times 10^{-10}$, $q=1$, $e = N/3$. Also, the $Q_{max}$ will be equal to QBER. 

 Using Eqn. \ref{eqn:finite_key_len} and the values discussed above we can determine that in order to be secret, the key length following privacy amplification ($\ell$) must be chosen to fulfil the inequality
 \begin{equation}
     \ell \leq  N(1-h_2(QBER +\mu))- N + k - \log{\frac{2}{\epsilon_{sec}^2 \epsilon_{cor}}}
 \end{equation}

\begin{figure}[t]
    \centering
    \includegraphics[width= 0.7\linewidth]{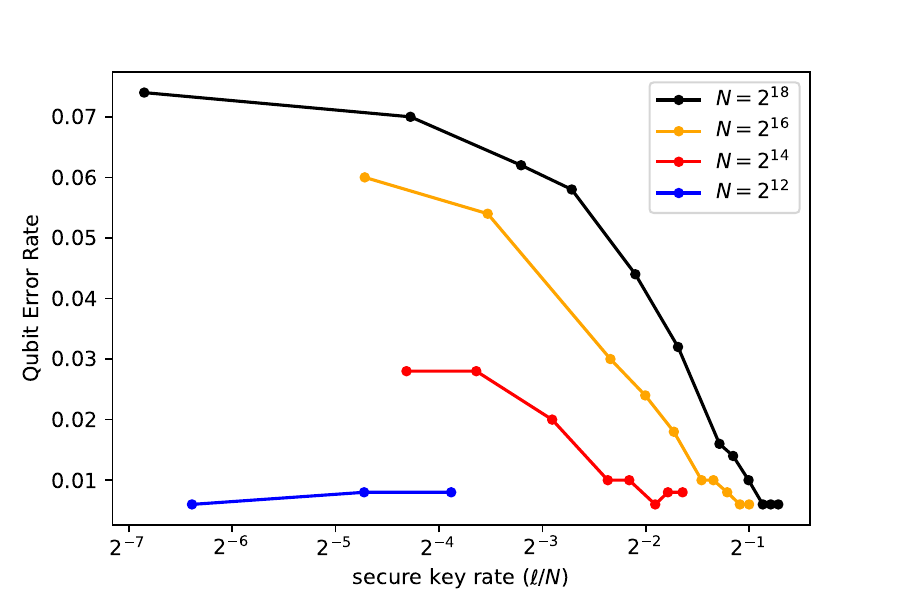}
    
    \caption{Variation of secure-key rate ($\frac{\ell}{N}$) with QBER $p$ (see Eqn. \ref{eqn:finite_key_len}). Here $\epsilon_{cor}= FER = 0.05$, $\epsilon_{sec}=0.5\times 10^{-10}$, $q=1$, and $e = N/3$}
    \label{fig:sec-key rate}
\end{figure}

According to the paper \cite{tomamichel2012tight} we plot the secret key rate $r = \frac{\ell}{N}$ shown in Fig. \ref{fig:sec-key rate}. This shows the secure-key rate established using our experiments and a two-universal hash function (see Def. \ref{def:two-universal-hash}) as an extractor. Note that, for block length $N=2^{10}$ no secret key could be extracted, therefore the data is not represented in Fig. \ref{fig:sec-key rate}.

 A detailed calculation of the secure key length including the relevant definitions can be found in \ref{appendix:Finite_key_analysis}.

\subsection{Infinite key analysis}
For an infinite key setting, using Eqn. \ref{eqn:finite_key_len}, the secret key rate can be reduced to (also see \cite{wolf2021quantum})
\begin{equation}
    r_{\infty} \geq 1- 2h(p)
\end{equation}

\section{Conclusion and future works}
\label{section:conclusion}
From the results shown in the paper, it can be concluded that code construction provided by Arikan using Bhattacharyya parameters \cite{arikan} produces good results when implemented on an underlying BSC. To implement such code construction we have provided an algorithm to calculate the \RS~ with a complexity of $\mathcal{O}(N\log N)$. The calculations of the Bhattacharyya parameter expressions for BSC are also shown. 
{ We note that such encoders can be designed independently of the decoders, and so can be flexibly used with different decoders. Our experiments, in particular experiment 2 (see sec. \ref{sec:expt2}), shows that for smaller block size ($N = 2^{10}$)   the \RS s of  BSC and BEC work similarly.  However, for larger block sizes ($N > 2^{10}$) the \RS~designed for BSC provides better error capacity, {\em i.e.} max. error corrected for a given code rate, for lower values of QBER and the difference increases with the increase in the block size ($N$). 

Our future work includes improving algorithm \ref{alg:rs} to obtain a more accurate estimate on the \RS, thereby bringing the experimental results closer to the Shannon limit. We note that Fig. \ref{fig:BEC-BSC comparison} c, is not aligned with Arikan’s suggestion that the code performance is insensitive to the choice of good channels. Verifying this result theoretically and with a wider set of experiments will the the subject of our future research directions.}

%\section*{References}
%\nocite{*}
%\bibliographystyle{iopart-num}
%\bibliography{references}

\bibliographystyle{plain}
\bibliography{references}

\appendix
\section{Proof of Proposition \ref{prop1} (Calculation of Bhattacharyya parameters)}\label{appendix:A}

The phenomenon of channel polarization takes $N$ independent copies of channel ($W$) as input and transforms them into polarized channels ($W_N^{(1)},...,W_N^{(N)}$). Arikan in his paper \cite{arikan} argues that this blockwise channel transformation can be broken into recursive single-step transformations. If we say that a single-step transformation produces two channels $W_{2}^{(1)}: \mathcal{X} \to \mathcal{\Tilde{Y}}$ and $W_{2}^{(2)}: \mathcal{X} \to \mathcal{\Tilde{Y}} \times \mathcal{X}$.    Then, a transformation ($W_N^{(i)},W_N^{(i)}$) $\to$ ($W_{2N}^{(2i-1)},W_{2N}^{(2i)}$) is possible only if there exists a channel ($W$) which maps $W: \mathcal{Y}^2 \to \mathcal{\Tilde{Y}}$, such that following is true (\cite{arikan} equation 22 and 23).   
\begin{align}
    W_{2N}^{(2i-1)}(y_1^{2N},u_1^{2i-2}|u_{2i-1}) = \sum_{u_{2i}}\frac{1}{2} W_N^{(i)}(y_1^N,u_{1,o}^{2i-2} \oplus u_{1,e}^{2i-2}|u_{2i-1} \oplus u_{2i})  \nonumber \\
    \cdot W_N^{(i)}(y_{N+1}^{2N}, u_{1,e}^{2i-2}|u_{2i})
\end{align}
\begin{align}
    W_{2N}^{(2i)}(y_1^{2N},u_1^{2i-1}|u_{2i}) = \frac{1}{2} W_N^{(i)}(y_1^N,u_{1,o}^{2i-2} \oplus u_{1,e}^{2i-2}|u_{2i-1} \oplus u_{2i}) \nonumber \\
    \cdot W_N^{(i)}(y_{N+1}^{2N}, u_{1,e}^{2i-2}|u_{2i})
\end{align}
where $u_{1,o}$ and $u_{1,e}$ are the odd and even input variables to the channel. Also here $u_is$ are input variables and to the channel and $y_is$ are the outputs. For simplicity let's make the following substitution:\\
$W \gets W_N^{(i)}$, $W' \gets W_{2N}^{(2i-1)}$, $W'' \gets W_{2N}^{(2i)}$, $u_1\gets u_{2i-1}$, $u_2\gets u_{2i}$, $y_1\gets (y_1^N,u_{1,o}^{2i-2}\oplus u_{1,e}^{2i-2})$, \\ $y_2\gets (y_1^N,u_{1,e}^{2i-2})$, and $f(y_1,y_2) \gets (y_1^{2n},u_1^{2i-2})$. 

Bhattacharyya parameter for a general B-DMC can be written as:
\begin{equation}
    Z(W) = \sum_{y \in \mathcal{Y}}\sqrt{W(y|0)\cdot W(y|1)}
\end{equation}
which for BSC becomes:
\begin{equation}
    Z(W) = 2\sqrt{(1-p)\cdot p}
\end{equation}

From s A.1 and A.3, we can write the Bhattacharyya parameter expression for the transformed channel ($W'$) as:
    \begin{align*}
        Z(W') &= \sum_{y_1^2}\sqrt{W'(f(y_1,y_2)|0)\cdot W'(f(y_1,y_2)|1)}\\
        &=\sum_{y_1^2}\frac{1}{2}\sqrt{ W(y_1|0)W(y_2|0)+ W(y_1|1)W(y_2|1)}\\
        &\cdot\sqrt{W(y_1|0)W(y_2|1)+ W(y_1|0)W(y_2|1)}
    \end{align*}

For BSC we can write:
\begin{align*}
    &W(0|0) = W(1|1) = 1-p, \\
    &W(1|0) = W(0|1) = p
\end{align*}
On taking all the possible cases of output ($y_1 = 0$ and $1$ $\&$ $y_2 = 0$ and $1$)

    \begin{align*}
     Z(W') &=\frac{1}{2} \sqrt{ W(0|0)W(0|0)+ W(0|1)W(0|1)}\cdot\sqrt{W(0|0)W(0|1)+ W(0|0)W(0|1)}\\
        &+\frac{1}{2} \sqrt{ W(0|0)W(1|0)+          W(0|1)W(1|1)}\cdot\sqrt{W(0|0)W(1|1)+ W(0|0)W(1|1)}\\
        &+\frac{1}{2} \sqrt{ W(1|0)W(0|0)+ W(1|1)W(0|1)}\cdot\sqrt{W(1|0)W(0|1)+ W(1|0)W(0|1)}\\
        &+\frac{1}{2} \sqrt{ W(1|0)W(1|0)+ W(1|1)W(1|1)}\cdot\sqrt{W(1|0)W(1|1)+ W(1|0)W(1|1)}\\
        &= \frac{1}{2} \sqrt{((1-p)^2+p^2)(2p(1-p))}+ \frac{1}{2} \sqrt{((1-p)^2+p^2)(2p(1-p))}\\
        &= 2\sqrt{2} \sqrt{p(1-p)}\cdot \sqrt{p^2+(1-p)^2}
    \end{align*}

Using equation A.4 in the previous equation. We get:
\begin{equation}
    Z(W') = Z(W)\sqrt{2-{Z(W)}^2}
\end{equation}

Similarly, the Bhattacharyya parameter expression for the  transformed channel ($W''$) can be written with the help of equation A.2 as follows: 

\begin{align*}
    Z(W'') &=\sum_{y_1^2,u_1}\sqrt{W''(f(y_1,y_2),u_1|0)\cdot W''(f(y_1,y_2),u_1|1)}\\
    &= \sum_{y_1^2,u_1}\frac{1}{2} \sqrt{W(y_1|u_1)W(y_2|0)}\cdot \sqrt{W(y_1|u_1\oplus1)W(y_2|1)}\\
    &=\sum_{y_2} \sqrt{W(y_2|0)W(y_2|1)}\times \sum_{u_1}\frac{1}{2}\sum_{y_1}{W(y_1|u_1)W(y_1|u_1\oplus1)}\\
    &=Z(W)\cdot Z(W)
\end{align*}
Therefore,
\begin{equation}
    Z(W'') = Z(W)^2
\end{equation}
Rewriting the substituted variables in terms of the original variables. The Bhattacharyya parameter for polarized BSC can be written as:
\begin{align*}
    &Z(W_{2N}^{(2i-1)}) = Z(W_{N}^{(i)})\sqrt{2-{Z(W_{N}^{(i)})}^2} \\
   &Z(W_{2N}^{(2i)}) = Z(W_{N}^{(i)})^2.
\end{align*}

\section{Finite Key Analysis}
\label{appendix:Finite_key_analysis}
\subsection{Definitions}
To understand the security definitions we need to understand what systems and states are involved in this procedure. Let $X$ be a random variable that describes Alice's bit string and $x \in X$ are its realizations following probability distribution $p_X(x)$. Here we can encode $\{\ket{x}\}$ as a collection of an orthonormal set of some Hilbert space $\mathcal{H}_X$. Let $\rho_E^x$ be the system of Eve correlated with Alice corresponding to realizations $x \in X$. In this case, we can define a classical-quantum ensemble as:
\begin{equation}
    \rho_{XE} = \sum_{x \in X} p_X(x) \rho_E^x \otimes \ket{x}\bra{x}
\end{equation}
We will use the classical-quantum ensemble to prove the security of our scheme.

\begin{definition}[Quantum conditional min-entropy \cite{wolf2021quantum}]
The quantum conditional min-entropy of a joint state of Alice and Bob $\rho_{A,B}$ is defined as:
\begin{equation}
    H_{min} (A|B) = -\log \min_{\sigma_{B}}\min_{\lambda} {\{\lambda: \rho_{AB} \leq \lambda \cdot \mathbb{1} \otimes \sigma_B}\}
\end{equation}
where the minimum is over all states $\sigma_B$.
\end{definition}
We will use this definition to define smooth conditional min-entropy.

\begin{definition}[Smooth conditional min-entropy \cite{renner2008security, wolf2021quantum}]
\label{def:smooth-min}For a classical quantum state $\rho_{XE}$ the smooth conditional min-entropy is defined by
\begin{equation}
    H_{min}^{\epsilon}(X|E) = \max_{\rho' \in {\cal B}^{\epsilon}(\rho) } H_{min} (X|E)_{\rho'}
\end{equation}where, ${\cal B}^{\epsilon}(\rho)$ is a closed ball with radius $\epsilon$ and center $\rho$.
\end{definition}
This is the amount of uniform randomness Alice can extract from her classical random variable $X$ which is correlated with Eve's system.

\begin{definition}[($\delta, \epsilon$)- strong quantum proof randomness extractor \cite{konig2011sampling}]
\label{def:quantum-extractor}
It is a function $\ext: \{0,1\}^n \times \{0,1\}^d \to \{0,1\}^m$ if for all classical-quantum states $\rho_{XE}$ with a classical random variable $X \in \{0,1\}^n$ with min-entropy $H_{min}(X|E) \geq \delta$ and a uniform random seed $Y \in \{0,1\}^d$ we have
\begin{equation}
    \frac{1}{2} \norm{\rho_{Ext(X,Y)YE} - \frac{\mathbb{1}}{2^m} \otimes \rho_Y \otimes \rho_E} \leq \epsilon
\end{equation}
    Here, $\norm{\rho}$ is the trace distance of $|\rho| = \sqrt{\rho ^ \dag \rho}$, and $\frac{\mathbb{1}}{2^m} \otimes \rho_Y \otimes \rho_E$ is the ideal situation. 
\end{definition}
This extractor can be used to extract randomness from a classical system correlated with a quantum system.
\begin{definition}[Two-universal hash function (2-UHF)]
\label{def:two-universal-hash}
Let $\mathcal{F}$ be a family of functions from an alphabet 
$\chi \to \mathcal{Z}$. $p_{\mathcal{F}}$ is a probability distribution on $\mathcal{F}$. The two-universal hash function is defined by the pair $(\mathcal{F}, p_{\mathcal{F}})$ such that,
\begin{equation}
    Pr_{f\in {\mathcal{F}}}[f(x) = f(x')] \leq \frac{1}{|Z|}
\end{equation}
for any $x, x' \in \chi$ with $x \neq x'$. $f \in \mathcal{F}$.
\end{definition}
We can say that for $0\leq\ell\leq n$, there exist a 2-UHF from $\{0,1\}^n$ to $\{0,1\}^{\ell}$ (see \cite{renner2008security} lemma 5.4.2). We use this as a quantum-proof randomness extractor which further leads to the definition of quantum leftover hash lemma.

\begin{lemma}[Quantum Leftover Hash Lemma \cite{konig2011sampling}]
\label{lemma:QLHL}
Let $\rho_{f_{PA}(K_A)YE}$ be the state after applying a random two-universal hash function $f_{PA}$ to Alice's raw key $K_A$. Then for every $\epsilon^{\prime}>0$,
\begin{equation}
 \frac{1}{2} \norm{\rho_{f_{PA}(K_A)YE} - \rho_{U} \otimes \rho_{YE}} = D(\rho_{f_{PA}(K_A)YE}, \rho_{U} \otimes \rho_{YE}) \leq 2\epsilon^{\prime}+ \frac{1}{2} \sqrt{2^{\ell - H_{min}^{\epsilon'}(K_A|E)}}    
\end{equation}
Here $\rho_U = \sum_{u\in Z} \frac{1}{Z} \ket{u}\bra{u}$ is the maximally mixed state over the key space.    
\end{lemma}
This helps us to find the secret key length for the finite case (see Sec. \ref{sec:finite-key-analysis}).

\subsection{Finite key analysis: Calculations}
\label{sec:finite-key-analysis}

 In this section, we will calculate the expression of the secure key length for our protocol (see Fig. \ref{fig:full_protocol}), based on the work of \cite{tomamichel2012tight}. The main challenge is to find a lower bound of the min-entropy so that we can extract randomness by using the $(\delta,\epsilon)$ quantum proof randomness extractor (see Def. \ref{def:quantum-extractor}). The steps for classical post-processing for finite key analysis are discussed as follows:
 \begin{enumerate}
     \item In parameter estimation step where QBER is estimated, we fix the maximum tolerated error rate as $Q_{max}$ for QBER.
     \item In the error-correction phase, the leakage due to the involvement of the classical public channel is leak$_{EC}$. After this phase to verify the correctness of the error correction step, Alice and Bob use a two-universal hash function (see Def. \ref{def:two-universal-hash}) $f_{EC}$. If $|\mathcal{Z}| = 2^{\lceil \log{\frac{1}{\epsilon_{cor}}} \rceil}$, then we can verify the protocol is $\epsilon_{cor}$-correct \cite{wolf2021quantum}. Therefore the total leakage after this step is, 
     \begin{equation}
         leak_{EC} + \lceil \log{\frac{1}{\epsilon_{cor}}} \rceil \leq leak_{EC} + \log{\frac{2}{\epsilon_{cor}}}
     \end{equation}
     \item In the privacy amplification phase, Alice and Bob will use another two-universal hash function $f_{PA}$ as a quantum-proof randomness extractor \cite{renner2008security}. Using quantum leftover hash lemma (see Lemma \ref{lemma:QLHL}) and the previously calculated leakage, we can find the secret key length ($\ell$) as \cite{tomamichel2012tight}
     \begin{equation}
         \ell \leq H_{min}^{\epsilon'}(K_A|E) - leak_{EC} - \log{\frac{2}{\epsilon_{sec}^2 \epsilon_{cor}}} 
     \end{equation}
 \end{enumerate}

According to the paper \cite{tomamichel2012tight}, the min-entropy $H_{min}^{\epsilon'}(K_A|E)$ can be lower bounded by 
\begin{equation}
    H^{\epsilon'}_{min}(K_A|E) \ge N(q-h_2(Q_{max} +\mu))
\end{equation}
where, $\mu = \sqrt{\frac{(e+1)(N+e)}{e^2 N} \log\frac{1}{\epsilon'}}$, and $q = \log \frac{1}{c}$, $c$ is incompatibility between two measurements. For the case of BB84, $q=1$. 

Let $E'$ include all the quantum and classical information available to Eve. Therefore we can write by chain rule of smooth min-entropy \cite{renner2008security}
\begin{align}
    H_{min}^{\epsilon'}(K_A|E') &\geq H_{min}^{\epsilon'}(K_A|E)- leak_{EC} - \log{\frac{2}{\epsilon_{cor}}}\nonumber\\
    &\geq N(q-h_2(Q_{max} +\mu))- leak_{EC} - \log{\frac{2}{\epsilon_{cor}}}
\end{align}
using this along with the quantum leftover hash lemma (see lemma \ref{lemma:QLHL}), we can write. 
\begin{align}
    \frac{1}{2} \norm{\rho_{K_{AE'}}-\rho_U \otimes \rho_E'} &\leq 2\epsilon'+ \frac{1}{2} \sqrt{2^{\ell -H_{min}^{\epsilon'}(K_A|E)+ leak_{EC} + \log{\frac{2}{\epsilon_{cor}}}}}\nonumber\\
    &\leq 2\epsilon'+ \frac{1}{2} \sqrt{2^{\log(\epsilon_{sec})^2}}\nonumber\\
    &\leq 2\epsilon' +\frac{\epsilon_{sec}}{2} 
\end{align}
In our case $\epsilon' = \frac{\epsilon_{sec}}{4}$. Therefore,
\begin{equation}
    \frac{1}{2} \norm{\rho_{K_{AE'}}-\rho_U \otimes \rho_{E'}} \leq \epsilon_{sec}
\end{equation}
which proves the secrecy of our protocol as per definition (see Eqn. \ref{eqn:secofQKD}).

So, we can write the secret key length  \cite{tomamichel2012tight} as
\begin{equation}
\label{eqn:finite_key_len1}
    \ell \leq  N(q-h_2(Q_{max} +\mu))- leak_{EC} - \log{\frac{2}{\epsilon_{sec}^2 \epsilon_{cor}}}
\end{equation}

\section{Nakassis \& Mink's key establishment protocol \cite{nakassis2014}}\label{appendix:B}

Based on the polar encoding procedure discussed in sec. \ref{subsec:3.2}, Nakassis and Mink \cite{nakassis2014} proposed a complete protocol for key-establishment
as follows:\\[-20 pt]

\begin{figure*}[ht]
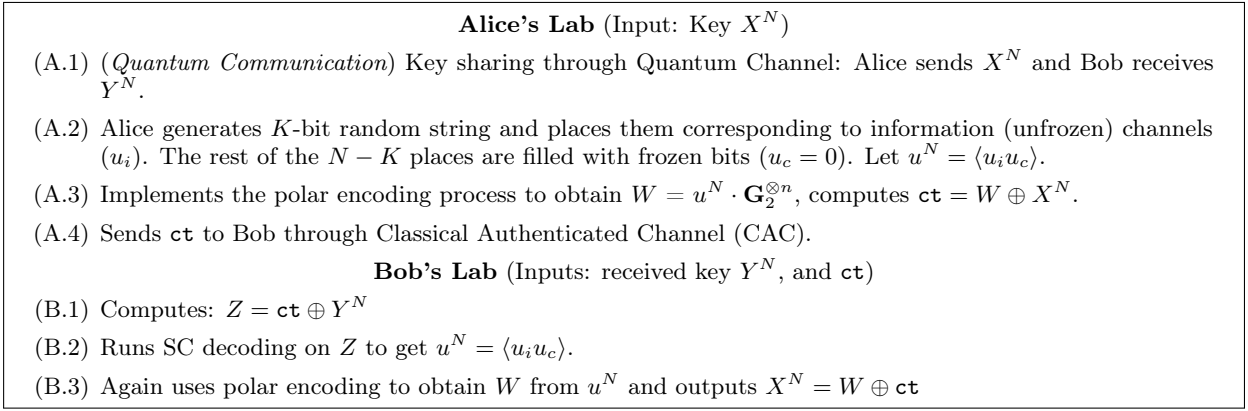

    \hspace*{2mm}\framebox[\textwidth][c]{
			\begin{minipage}{0.95\textwidth}

{\bf Alice's Lab} (Input: Key $X^N$)
\begin{itemize}

\item[(A.1)] ({\em Quantum Communication}) Key sharing through Quantum Channel: Alice sends $X^N$ and Bob receives $Y^N$.
\item[(A.2)] Alice generates $K$-bit random string and places them corresponding to information (unfrozen) channels ($u_i$). The rest of the $N-K$ places are filled with frozen bits ($u_c=0$). Let $u^N= \langle u_i u_c \rangle$.
\item[(A.3)] Implements the polar encoding process to obtain $W$ = $u^N \cdot \mathbf{G}_2^{\otimes n}$, computes
    ${\tt ct} = W \oplus X^N$.
\item[(A.4)] Sends ${\tt ct}$ to Bob through Classical Authenticated Channel (CAC).
\end{itemize}
{\bf Bob's Lab} (Inputs: received key $Y^N$, and ${\tt ct}$)
\begin{itemize}
\item[(B.1)] Computes: $Z = {\tt ct} \oplus Y^N$
\item[(B.2)] Runs SC decoding on $Z$ to get $u^N =  \langle u_i u_c \rangle$.
\item[(B.3)] Again uses polar encoding to obtain $W$ from $u^N$ and outputs $X^N=W \oplus {\tt ct}$
\end{itemize}

   \end{minipage}
}
   
    \caption{Nakassis $\&$ Mink's key-exchange protocol}
    \label{fig:nakassis_protocol}
\end{figure*}

\end{document}